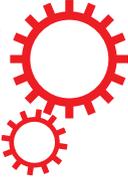

# SCIENTIFIC REPORTS

**OPEN**

# Enhancing laser-driven proton acceleration by using micro-pillar arrays at high drive energy

Dimitri Khaghani[1,2], Mathieu Lobet[3,4], Björn Borm[1,2], Loïc Burr[5,6], Felix Gärtner[1,2], Laurent Gremillet[3], Liana Movsesyan[5,6], Olga Rosmej[1], Maria Eugenia Toimil-Molares[5], Florian Wagner[1,7] & Paul Neumayer[1]

The interaction of micro- and nano-structured target surfaces with high-power laser pulses is being widely investigated for its unprecedented absorption efficiency. We have developed vertically aligned metallic micro-pillar arrays for laser-driven proton acceleration experiments. We demonstrate that such targets help strengthen interaction mechanisms when irradiated with high-energy-class laser pulses of intensities ~$10^{17-18}$ W/cm². In comparison with standard planar targets, we witness strongly enhanced hot-electron production and proton acceleration both in terms of maximum energies and particle numbers. Supporting our experimental results, two-dimensional particle-in-cell simulations show an increase in laser energy conversion into hot electrons, leading to stronger acceleration fields. This opens a window of opportunity for further improvements of laser-driven ion acceleration systems.

Laser-driven proton beams have attracted increasing interest due to their unique properties and many applications[1,2]. Over the past 15 years, experiments have shown proton beams with energies up to several tens of MeV[3,4], accelerated from the rear surface of thin planar foils irradiated at laser intensities exceeding $10^{18}$ W/cm² via a mechanism known as target normal sheath acceleration (TNSA)[5]. Large experimental and theoretical effort has been devoted to optimizing the ion beam parameters so as to demonstrate the suitability of this all-optical technique for many purposes, ranging from tumor therapy[6] to proton radiography[7] and the creation of warm dense matter[8]. Improved laser performances in terms of pulse energy, intensity and temporal profile have enabled major advances in increasing the total yield and maximum energy of TNSA beams. This progress mainly results from enhanced energy conversion into the hot electrons that set up the sheath electric field responsible for the ion acceleration.

Another promising way to improve ion acceleration is to use tailored-surface targets due to the variety of beneficial physical processes that they may trigger[9]. Thus, for example, their larger interaction surface leads to an increase in laser absorption. Hollow surfaces, such as nano-hole targets, can even act as light-traps[10]. In these cases, the laser pulse is not directly reflected at the overcritical target front surface, but interacts for an extended time in the inner volume of the target. Moreover, pointy and needle-like structures present a topographic advantage, provided by the so-called mass-limited effect[11]. The return current induced by bulk electrons compensating for charge losses as hot electrons are being pulled away by the laser field, is confined inside the elongated structures. These anisotropic particle flows tend to enhance the electrostatic fields within the target structure, leading eventually to the acceleration of ions by Coulomb explosion. Furthermore, structured target surfaces can also benefit from specific laser-matter interaction mechanisms requiring peculiar incident geometries, such as the Brunel effect[12], which leads to a significant increase in the efficiency of hot electron generation.

Few examples of proton acceleration with tailored target geometries have been reported. For instance, polyethylene terephthalate foils covered with polystyrene nano-spheres and irradiated at a laser intensity of $5 \times 10^{19}$ W/cm² (1 J in 30 fs) led to a 60% increase in cut-off proton energy, i.e. from 5.3 to 8.6 MeV[13]. In the same experiment, the MeV-proton yield underwent a 5-fold rise. Thin grating targets demonstrated a 2.5-fold increase in proton cut-off energy[14]. It has also been reported that samples contaminated by a layer of E. Coli bacteria can significantly enhance the hot-electron production and, consequently, generate stronger accelerating sheath fields[15, 16].

[1]GSI Helmholtzzentrum für Schwerionenforschung, Plasma Physics, D-64291, Darmstadt, Germany. [2]Johann Wolfgang Goethe-Universität, Institute of Applied Physics, D-60438, Frankfurt am Main, Germany. [3]CEA, DAM–DIF, F-91297, Arpajon, France. [4]CELIA, UMR 5107, Université de Bordeaux – CNRS – CEA, F-33405, Talence, France. [5]GSI Helmholtzzentrum für Schwerionenforschung, Materials Research, D-64291, Darmstadt, Germany. [6]Technische Universität Darmstadt, Materials and Earth Sciences, D-64287, Darmstadt, Germany. [7]Helmholtz-Institut Jena, D-07743, Jena, Germany. Correspondence and requests for materials should be addressed to D.K. (email: d.khaghani@gsi.de) or P.N. (email: p.neumayer@gsi.de)

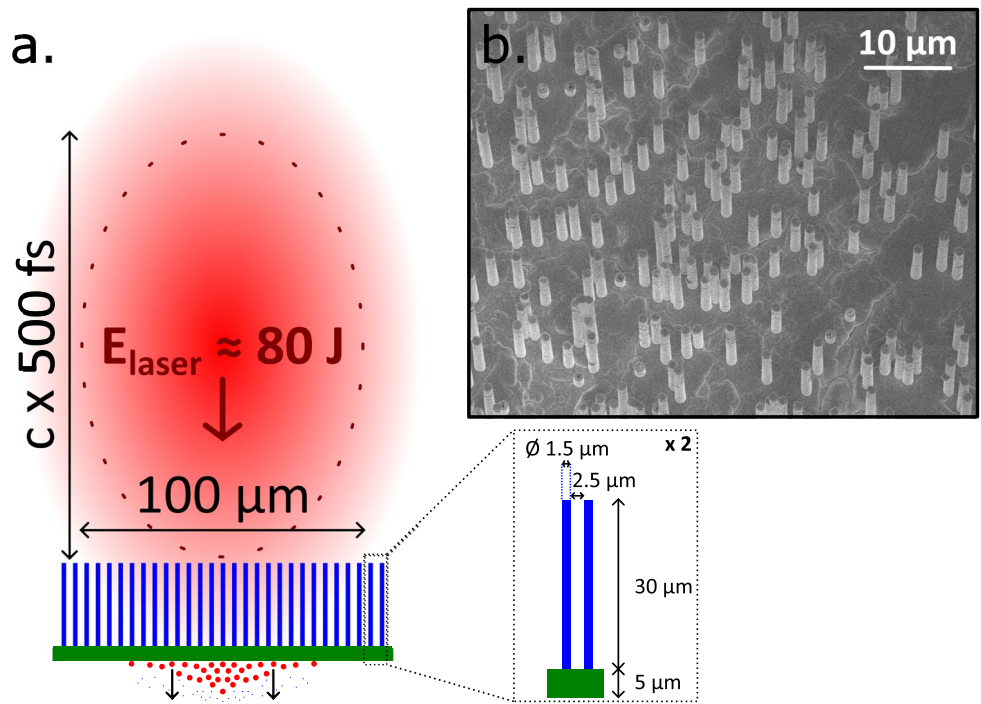

**Figure 1.** (**a**) A pulse from the PHELIX laser (red) irradiates a micro-pillar array (blue), inducing ion acceleration from the rear side of the film layer (green). All dimensions are to scale. Scanning electron microscopy (SEM) is used to characterize the targets before laser irradiation. (**b**) It provides information concerning the dimensions of a representative array target consisting of vertically aligned Cu-pillars with diameter 1.5 μm and length 30 μm standing on a Ag layer (the target is observed under an angle of ~6° from the normal to the support layer).

Our study aims at taking advantage of high-aspect-ratio micro-structures in order to deposit energy deeper into the target, unlike all aforementioned targets which rapidly produce a plasma above the critical density, $n_c \approx 10^{21} \lambda_{\mu m}^{-2}$ cm$^{-3}$ (where $\lambda_{\mu m}$ is the laser wavelength) at their front surface. For example, the irradiation of Ni nano-wire arrays by ultra-short sub-joule laser pulses ($I \sim 5 \times 10^{18}$ W/cm$^2$) recently gave rise to enhanced x-ray emission and high-energy-density conditions[17]. These results were attributed to the increased laser absorption allowed by the high aspect ratio and hollow architecture of those samples.

Further topographic advantages are expected from such wire arrays. For instance, vertical wires should favor Brunel-type electron heating, as the electrons located at their wall can directly interact with the locally $p-$ polarized laser field, which pulls them into the vacuum and accelerates them within a few laser cycles before releasing them at high velocity[18, 19]. The role of these relativistic electrons may be considerable, as they have a mean free path of a few 100 μm, thus filling the whole volume of the target. It has also been reported both from simulations and experiments that electrons can be guided along nano-structures in brush-like targets[20]. The influence of nano-structured surfaces on the hot electron distribution has also been investigated by modeling ultra-intense laser-matter interactions with the help of a particle-in-cell (PIC) code[21], and Monte-Carlo simulations predicted an enhancement of Bremsstrahlung emission[22]. More recently, arrays of Si micro-wires irradiated by ultra-short joule-class laser pulses ($I \sim 1 \times 10^{21}$ W/cm$^2$) have generated electron beams with increased total and cut-off energies, from 30 to 70 MeV[23]. Furthermore, it has also been demonstrated that aligned Cu-nanorod arrays irradiated by sub-joule ultra-short laser pulses ($I \sim 3.5 \times 10^{18}$ W/cm$^2$) can generate intense broadband terahertz sources[24].

## Results

In this work, we report on the strong enhancement of proton acceleration from μm-thin metallic foils covered with vertically aligned free-standing μm-thick metallic wires, also known as micro-pillar arrays, irradiated by high-energy (~100 J), ultra-high-contrast, sub-picosecond laser pulses delivered by the PHELIX system (Fig. 1). Our prime motivation was to demonstrate the potential of wire array targets for increasing the laser coupling efficiency at high drive energy. This represents a double challenge, both experimentally and numerically. Indeed, considering the current state-of-the-art laser technology, high-energy intense laser pulses have a rather long duration (>100 fs), which constrains the target dimensions, requires a laser contrast of ultra-high level and is highly computationally expensive for numerical kinetic simulations. We produced μm-thick pillars with the largest possible aspect ratio allowed by our production method (30 μm in length), and an average spacing of a couple of microns so that the target remains clearly over-critical. Hot-electron as well as proton acceleration were investigated, evidencing a strong increase in both particle yields and cut-off energies. The emission of TNSA carbon ions was also revealed. These results are supported with state-of-the-art PIC simulations which demonstrate that micro-pillar arrays yield a large enhancement in hot-electron production, which boosts the subsequent TNSA process.

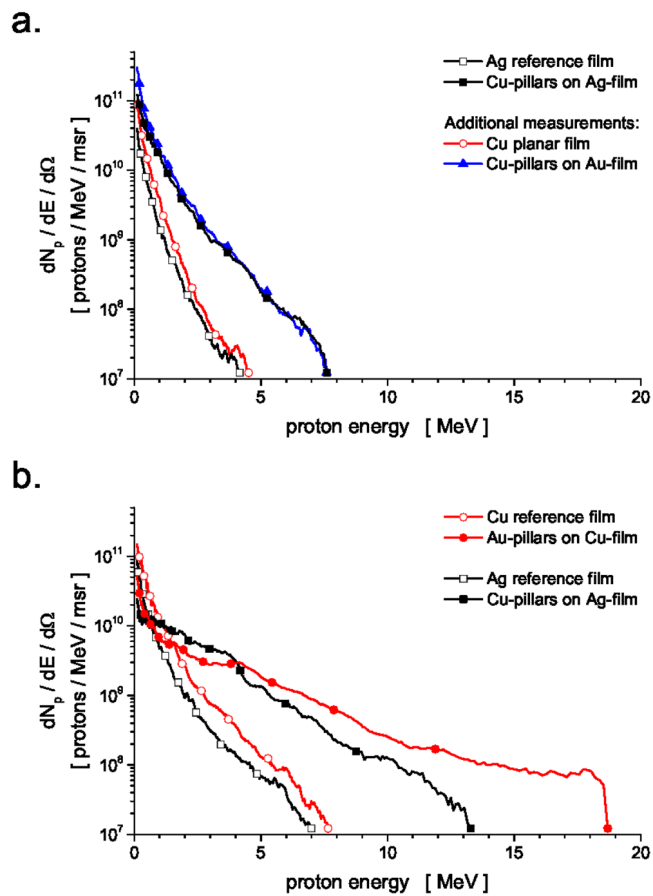

**Figure 2.** The energy distribution of protons accelerated from the rear side of micro-pillar targets (solid symbols) was measured and compared with those from flat targets (empty symbols), irradiated at a laser intensity of (**a**) $2 \times 10^{17}$ W/cm$^2$ and (**b**) $2 \times 10^{18}$ W/cm$^2$. Different planar film materials were used: copper (red), silver (black) and gold (blue).

**Experimental study.** At a rather low laser intensity of $2 \times 10^{17}$ W/cm$^2$ (Fig. 2a), a cut-off proton energy of ~4 MeV was recorded for a Ag reference film, and twice this value (~8 MeV) for Cu micro-pillar arrays standing on a Ag planar film. Integrating each spectrum for energies larger than 3 MeV, we obtain an estimate for the number of protons per unit solid angle in the beam, namely $4 \times 10^7$ protons/msr and $1.2 \times 10^9$ protons/msr for planar films and Cu-pillar samples, respectively. Hence, the proton yield was increased 30-fold using micro-pillar arrays. As also shown in Fig. 2a, we conducted additional measurements using a Cu planar film that demonstrated similar results to those from the Ag reference film, and Cu micro-pillar arrays standing on a Au film which performed comparably to the Cu-pillars on a Ag-film.

Increasing the laser intensity to $2 \times 10^{18}$ W/cm$^2$ (Fig. 2b) resulted in cut-off proton energies of ~7 and ~8 MeV for Ag and Cu planar films, respectively. Consistently with the lower-intensity case, the spectra of protons accelerated from Au-pillar targets exhibit a mean cut-off energy of ~19.5 MeV, i.e. an increase by a factor of 2.4. Targets with Cu-pillars yielded an increase in cut-off energy by 85% (up to ~13 MeV), compared to pillar-free planar films. In terms of particle number (for protons with energy >3 MeV), we measured $4 \times 10^8$ and $9 \times 10^8$ protons/msr for Ag and Cu planar films, respectively. In comparison, the micro-pillar arrays exhibited a 20-fold increase in proton yield with $8 \times 10^9$ and $1.5 \times 10^{10}$ protons/msr for Cu-pillar and Au-pillar targets, respectively.

In complement to the proton measurements, an electron spectrometer was used to record the energy distribution of the electrons escaping from the front side of the target. The results obtained when applying a laser intensity of $2 \times 10^{17}$ W/cm$^2$ are displayed in Fig. 3. The high-energy tail of the electron distribution from a planar Ag reference target can be fitted to a Boltzmann distribution with a temperature of ~140 keV, consistent with the so-called Beg's scaling law[25] and in agreement with previous measurements performed under similar conditions[26]. We measured an integrated flux of $6 \times 10^6$ electrons/msr above 500 keV. By contrast, the three different types of micro-pillar arrays consistently outperformed the planar target by yielding a 4-fold increase in the number of hot electrons ($2.5 \times 10^7$ electrons/msr) and a 2-fold increase in their temperature (~290 keV).

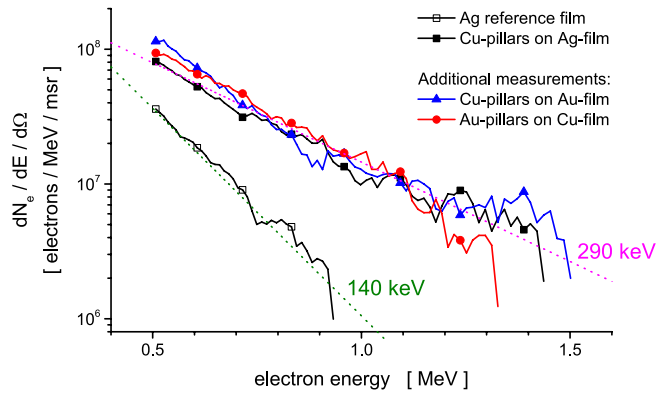

**Figure 3.** The energy distribution of electrons escaping from micro-pillar targets (solid symbols) irradiated at $2 \times 10^{17}$ W/cm$^2$ was measured and compared with those from a flat target (empty symbols). Different planar film materials were used: copper (red), silver (black) and gold (blue). The experimental data were fitted to a Boltzmann distribution (dashed lines).

**Particle-in-cell simulation.** 2D-PIC simulations were performed with the code calder[27] for two different laser strength parameters $a_o = 0.24$ ($I \sim 1 \times 10^{17}$ W/cm$^2$) and $a_o = 1.2$ ($I \sim 2 \times 10^{18}$ W/cm$^2$), where $a_o$ is the peak value of the normalized vector potential $eA/m_e c^2$. We first illustrate the capability of the laser to propagate through the micro-pillar array (Fig. 4a and b). Figure 4c shows that, for $a_o = 0.24$, the gaps between adjacent micro-pillars are essentially free of electrons due to their moderate heating. Under these conditions, the laser field can propagate. For $a_o = 1.2$ (Fig. 4d), the energetic electrons expelled from the pillar walls fill in the interstice spaces with a density ~0.1 $n_c$ (where $n_c$ is the critical density). The laser field is thus still able to propagate through the array, despite perturbations caused by the electrons. Overall, for both laser intensity values the conditions for laser propagation are fulfilled and, as expected, electrons heated up to higher temperatures upon a stronger laser field can spread over longer distances in the gaps.

We restrict the subsequent analysis to those electrons (i.e. electrons with energies >100 keV) able to reach the rear side of the target, and hence mainly contribute to ion TNSA. The resulting conversion efficiencies were estimated by computing the ratio of the hot-electron energy gain and the input laser energy. For $a_o = 1.2$, the simulation shows an increase in coupling of laser energy into hot electrons by a factor of 3.8, comparing the Cu micro-pillar array target to a Cu planar foil without pillars (14% and 3.7% conversion efficiency, respectively). Remarkably enough, the coupling efficiency into >1 MeV electrons rises 30-fold (3.8% vs 0.13%) as a consequence of the laser pulse's penetrating through the interstices between the pillars, and irradiating an extended surface of the inner target structure. Additional interaction mechanisms (e.g. linked to the complex light pattern setting in as a consequence of multiple reflections off the pillar walls) may also occur and enhance the hot-electron production.

The reduced geometry and space-time scales considered for our simulations preclude a realistic description of TNSA at the rear side of the film layer. However, these simulations allow us to assess the potential of the pillar target for enhancing TNSA by providing information on the local hot electron properties. Figure 5 displays the spatial density distribution of electrons with energies >100 keV at three different times. At a time of 450 fs before the laser peak (Fig. 5a), the first hot electrons are being generated as the laser pulse reaches the tip of the pillar. Their high energy permits them to penetrate deep inside the pillar. When the laser intensity has increased to half its peak value (Fig. 5b), the tip of the pillar is filled with a relatively dense ($\sim n_c$) cloud of hot electrons.

Most of the hot electrons are then confined by strong charge-separation fields in the highly-ionized Cu pillar (at the end of the simulation, most of the ions are in a He- and H-like state, some are even fully ionized). The pulse is now irradiating the whole volume of the array, but it is not yet strong enough to generate a high density of hot electrons at the base of the pillar. Yet a few energetic electrons have already traveled across the supportive base film and set up the TNSA sheath field at its rear side. At the time of the laser peak (Fig. 5c), the whole pillar is filled with high-density (>$n_c$), energetic (~300 keV) electrons. At the rear side of the base film, the hot-electron temperature is twice higher for the pillar targets than for the pillar-free planar target (440 keV and 220 keV, respectively).

Concomitantly, the simulation follows the growth in the TNSA field, as shown step by step in Fig. 5. At early times, a nascent electric field is measured at the tip of the pillar. As the laser field penetrates deep inside the target, creating denser and hotter electrons, the TNSA field builds up at the rear side of the film and reaches values of the order of 0.1 TV/m. Figure 5d shows the acceleration sheath field in the vicinity of the film for pillar and pillar-free targets at the laser intensity peak. While the pillar target only yields a slight increase in the TNSA field strength, it enhances the associated electric potential more significantly, i.e. by ~25%, which should be considered as a lower estimate since the hot electron cloud has reached the boundary of the simulation domain. This gives an estimate of the maximum energy potentially gained by the protons. In addition, the propagation time of the laser pulse between the long pillars is extended. We think that thanks to the longer interaction time the TNSA field has more time to build up, and that the whole acceleration process may last longer.

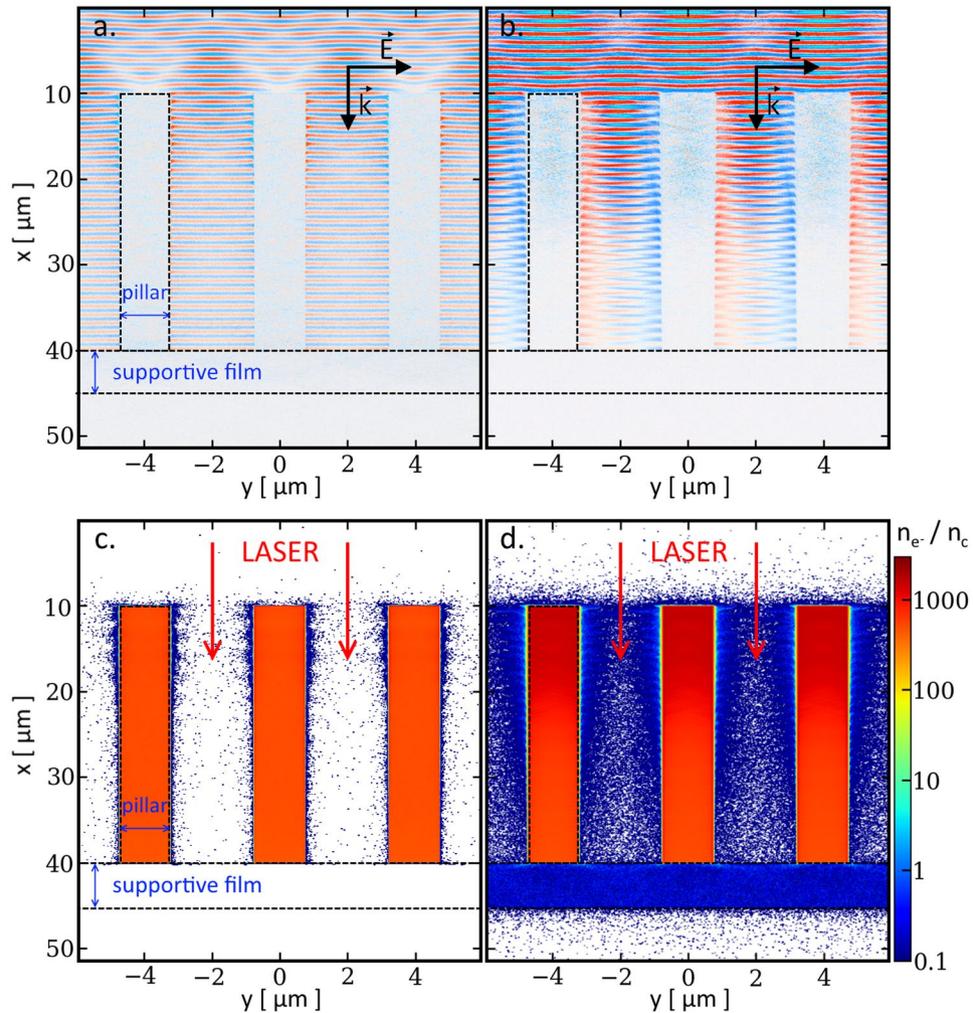

**Figure 4.** (**a**) and (**b**) The electric field of the laser propagates between the micro-pillars in the PIC simulation (field strength in arbitrary color scale). Electrons heated up by the laser expand into the interstices. (**c**) and (**d**) The numerical calculation provides a spatial distribution of the electron density ($n_{e^-}$), normalized to the critical density ($n_c$) value below which laser propagation is possible (dark blue regions). All plots correspond to the time when the laser intensity reaches half its maximum for two different laser strength parameters $a_o$ of 0.24 (left-hand panels) and 1.2 (right-hand panels).

## Discussion

Our simulations show that, even at the highest laser intensity applied (i.e. $\sim 2 \times 10^{18}$ W/cm$^2$), the electron distribution has not yet homogenized when the laser peak reaches the bottom of the pillar array. Indeed, most of the free electrons are trapped in the pillars, because the Coulomb potential is so deep that electrons cannot escape into the interstices. Only a few energetic ones (>1 MeV) are able to overcome this potential barrier and spread over the whole array. As this fraction of fast electrons is sparsely distributed, the formation of an over-critical cloud in the interstices is prevented, thus ensuring unhindered laser propagation. In other words, the geometrical advantages of the micro-pillar arrays are preserved as long as the full plasma, comprising both ions and over-critical bulk electrons, have not significantly expanded into the interstices. Achieving a better understanding of the kinetics of ion expansion in micro-structured geometries is thus crucial for the improvement of the TNSA proton acceleration systems.

Despite the ultra-high laser contrast level achieved during our experiments, the synthetic Gaussian pulse profile assumed in our simulations does not truly represent that of the experiment. We know from previous studies[28] that a low-intensity rising part of the laser pulse might preheat the target about 10 ps before the main pulse. Considering a worst-case estimate for this premature energy deposition and presuming that it is instantly and fully absorbed by the micro-pillar array 10 ps before the laser peak, we can estimate that the material would be preheated up to a few tens of eV. Assuming then transverse hydrodynamic expansion at the corresponding ion sound velocity, the plasma would expand over 10% to 20% of the inter-pillar distance in 10 ps, ensuring laser propagation in the remaining vacuum gaps. Moreover, upon higher laser intensities, our estimation suggests that ion expansion might weaken the beneficial effect of the micro-pillars. In that case, we would advise to increase the pillar spacing. Furthermore, still considering the ion sound velocity, one can assume that plasma expansion

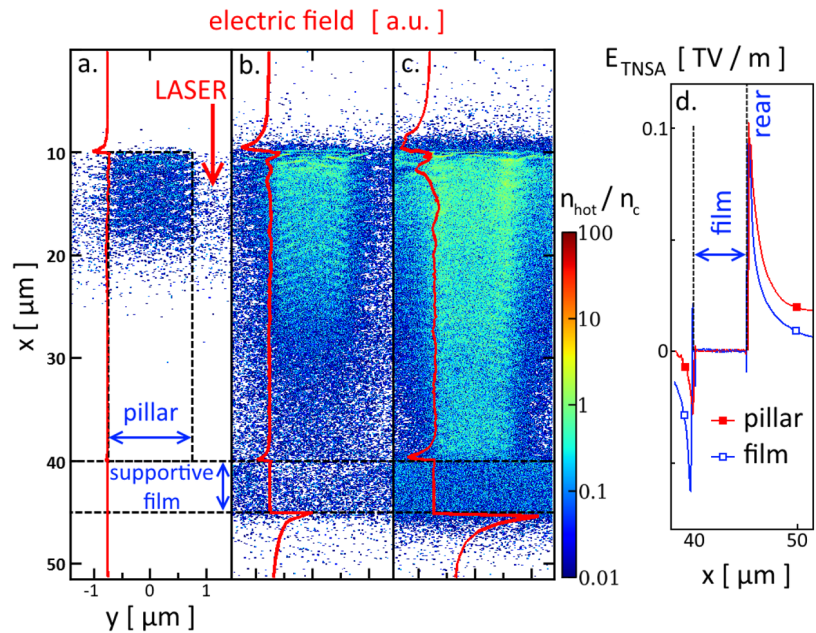

**Figure 5.** The PIC simulation follows the evolution of the spatial distribution of hot electrons ($E > 100$ keV) at (**a**) −450 fs, (**b**) −250 fs and (**c**) 0 fs with respect to the time when the pulse peak ($a_o = 1.2$) reaches the film layer. As the laser pulse moves forward into the pillar array, further energetic electrons are generated and an acceleration sheath field (red curve) builds up in the pillar direction. (**d**) At the time of the laser peak, this electric field extends differently behind the film for a planar target (blue empty squares) compared to a pillar array (red solid squares).

– prior or during the high-intensity interaction – is mainly governed by the element mass $m_i$, so that the Cu ions are expected to move $\sqrt{m_{Au}/m_{Cu}} \approx 2$ times faster than Au ions. As a result, Cu pillars may disintegrate earlier than their Au counterparts. This may explain the observed better TNSA performance of the Au micro-pillars compared to the Cu micro-pillars.

## Conclusion

We have experimentally demonstrated the potential of micro-pillar arrays with high aspect ratio and sufficient interspace to enhance proton acceleration by high-energy short-pulse lasers. We have found that micro-pillar arrays of Cu and Ag outperform all planar film targets, in terms of both cut-off energy and total proton yield (by factors of up to 2.4 and 30, respectively). Both experimental and simulation results demonstrate a significant increase in hot-electron production. This enhanced electron heating may stem from the widened effective interaction surface, as well as specific interaction mechanisms allowed by the target structure, such as Brunel heating. Overall, the enhancement of the hot-electron temperature and density leads to stronger ion-accelerating fields. Our experimental results and simulations provide important information on how to adjust the relevant geometrical target parameters, e.g. by increasing the inter-pillar distance or using high-$Z$ materials. The further development and optimization of micro-structured targets appears as a very promising avenue towards reaching even higher energies and particle numbers, which is important for applications such as proton radiography or laboratory astrophysics.

## Methods

**Target production.** Arrays of vertically aligned free-standing μm-size metallic wires, also called micro-pillar arrays in the main article (Fig. 1b), were produced by electro-deposition in etched ion-track membranes[29]. The etched ion-track membranes were produced by irradiation of polycarbonate foils (thickness 30 μm, Makrofol N, Bayer AG) with swift Au ions (11.1 MeV/u) at the UNILAC linear accelerator (GSI, Darmstadt). By chemical etching in a 6 M NaOH solution at 50 °C, each ion-track was selectively dissolved and enlarged into cylindrical channels with diameter 1.5 μm. Prior to etching, the irradiated foils were exposed to UV light (30 W, T-30 M, Vilber Lourmat) from both sides, each for 1 h.

Before growing the wire arrays, a supportive metallic film was placed onto one side of the polymer membrane so as to give mechanical stability to the latter target. It required sputtering a thin metallic layer (~100 nm), made of either Cu, Ag, or Au, that was then reinforced by depositing the properly speaking film of the same material. Cu-films were deposited using a copper-sulfate-based electrolyte (238 g/L $Cu_2SO_4$ and 21 g/L $H_2SO_4$) in a two-electrode configuration at room temperature applying a constant potential of −0.5 V vs a Cu anode. Ag-films were deposited using a silver-cyanide-based commercial electrolyte (Silver-BRIGHT-100, METAKEM GmbH) in a two-electrode configuration at room temperature applying a constant potential of −0.7 V vs a Ag anode. Au-films were deposited using a gold-sulphite-based commercial electrolyte (Gold-SF, METAKEM GmbH) in a two-electrode configuration at room temperature applying a constant potential of −0.8 V vs a Au anode.

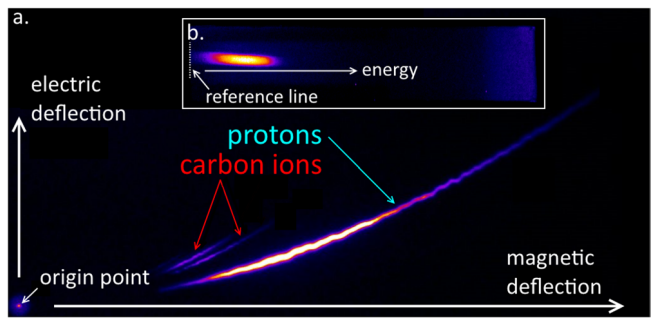

**Figure 6.** (**a**) The Thomson parabola ion spectrometer energy disperses and charge separates protons and carbon ions accelerated behind a micro-pillar target. (**b**) The electron spectrometer deflects the electrons escaping from the target according to their energies. The origin point and the reference line are fiducial markers for the energy calibration of the instruments.

To prepare the micro-wire arrays, after preparation of the supportive films, the metal of choice (Cu or Au) was electro-deposited inside the etched channels. In order to enable specific x-ray spectroscopy measurements, the fabrication processes were adjusted for each target so that supportive films and micro-wires consisted of two different materials, i.e. Cu-wires on Ag-film, Cu-wires on Au-film, Au-wires on Cu-film. Cu micro-wires were deposited using a copper-sulfate-based electrolyte (238 g/L $Cu_2SO_4$ and 21 g/L $H_2SO_4$) in a two-electrode configuration at 60 °C applying a constant potential of −0.1 V vs a Cu anode. Au micro-wires were deposited using a gold-cyanide-based electrolyte (50 mM $KAu(CN)_2$ and 0.25 M $Na_2CO_3$) in a three-electrode configuration at 60 °C applying a constant potential of −1.1 V vs a Ag/AgCl reference electrode (a Pt wire served as counter electrode). The potentiostatic growth process was monitored via current-vs-time measurements. This allowed stopping the deposition process as soon as the metal had filled the 30 μm-long channels, and before the material grew outside the channels. Finally, the achieved samples were revealed by dissolution of the polymer membrane in a dicholoromethane solution and mounted on an aluminum ring for laser irradiation experiments.

The density of the arrays of micro-wires with length 30 μm and diameter 1.5 μm was in all cases $10^7$ wires. $cm^{−2}$, in order to yield an average density of 1/6 solid density or 80 times critical density. At the macroscopic scale the targets were over-dense, but at the microscopic scale they presented deep vacuum gaps, wider than the laser wavelength, allowing for laser penetration. The fabrication of the micro-wire arrays by electro-deposition in etched ion-track membranes ensures, unlike any other synthesis method, the suitability of our targets for TNSA applications thanks to the thin enough supportive film that lets hot electrons (with energy >100 keV) reach the relatively flat rear side of the sample and thus generate the acceleration sheath field. Although randomly distributed due to the seed ion beam, the micro-wires have accurate diameters and lengths which can be controlled independently by etching time and metal deposition time, respectively.

Last but not least, the planar reference films were systematically produced using the same method described above for the supportive films, except that the membranes were not irradiated with the ion beam, nor etched, and of course no wire arrays were grown.

**Experimental set-up.** The targets were irradiated at the Petawatt High-Energy Laser for heavy Ion eXperiments (PHELIX) in GSI, Darmstadt. The Nd:glass laser system delivered high-energy (up to 80 J on target) pulses at 1 μm wavelength and with a bandwidth-limited duration of 500 fs.

Previous experiments by other groups (see *Introduction*) conducted at ultra-short-pulse laser facilities were limited by the state-of-the-art laser technology to an energy of a few joules. Here, by contrast, we demonstrate the suitability of our targets for high-energy short-pulse laser irradiation. Consequently, new challenges emerged: the amplified spontaneous emission (ASE) level and ps-scale prepulses put our delicate targets at risk; also, the longer pulse duration becomes comparable with the closure time of the vacuum gaps due to the expanding preheated pillars. The first issue was addressed by improving the laser contrast quality. The second issue was overcome by reducing the laser intensity in order to slow down the hydrodynamic evolution of the target.

An ultra-fast optical parametric amplifier (uOPA) provided amplification by up to 5 orders of magnitude of the oscillator seed pulses. Parametric fluorescence was limited to the short (sub-ps) pump pulse duration, resulting in ultra-high-contrast pulses with an ASE level below $10^{−10}$ as measured in previous experiments[28]. In addition, we used a plasma mirror to boost the contrast ratio. It acts as an anti-reflective mirror (reflectivity R < 0.1%) as long as the incident laser intensity remains below the so-called ignition threshold ($I \sim 10^{12}$ W/$cm^2$). In this way, it reduces the early part of the laser pulse by 3 to 4 orders of magnitude. The reflectivity after ignition was measured at the PHELIX system[30], it is shown that it reaches a maximum of 90%. The laser beam was focused with an f/10 off-axis parabola. The laser intensity was limited to a range from $2 \times 10^{17}$ to $2 \times 10^{18}$ W/$cm^2$ by placing the target behind the focal plane so as to obtain laser spot sizes of 300 and 100 μm, respectively, thus irradiating ~7000 and ~800 wires.

We measured spectra of accelerated protons along the normal axis of the rear side of the target, using a Thomson parabola ion spectrometer. This diagnostic consists of a permanent magnet which energy-disperses charged particles, and two metallic plates set to high potential difference in order to obtain an electric field which spreads out the different ion species according to their charge-to-mass ratio. The ions were detected by means of Agfa MD-40 imaging plates, as shown on Fig. 6a where protons and carbon ions were recorded. Spectra of hot electrons escaping the front side of the target were recorded using a permanent magnet electron spectrometer with Fuji BAS-TR imaging plates, as shown on Fig. 6b. Unfortunately, we were unable to measure hot electron spectra in the high-intensity regime ($I \sim 2 \times 10^{18}$ W/cm$^2$) due to the smearing of the signal by a charge separation effect.

**Computational parameters.** To reduce the computational load due to the large experimental spatio-temporal scales, the simulation domain covers only a fraction of the transverse interaction region. The target then consists of three copper pillars of 30 μm length, 1.5 μm width, standing on a 5 μm-thick Al-layer (the choice of Al, of lower electron density than Cu or Ag, avoids the use of an excessively refined discretization). The pillars are periodically spaced with gaps of 2.5 μm width. The target is placed in a $12 \times 50$ μm$^2$ simulation box, extending over 10 μm in front and 6 μm behind the target. The domain contains a total of approximately $2.5 \times 10^8$ mesh points. At the beginning of the simulation, it is mostly empty (~68% of the domain), the pillars consist of solid density Cu$^{5+}$ ions (~22% of the domain), the rest (~10%) accounts for the Al-film. The boundary conditions are taken to be absorbing (resp. periodic) in the longitudinal (resp. transverse) direction for both particles and fields. The laser pulse is described with a 500 fs-FWHM Gaussian temporal profile for two different laser strength parameters ($a_o = 0.24$ and 1.2) The simulation runs over ~800 fs, resolved in 71,000 time steps. The laser peak in intensity reaches the film layer ~20 fs before the end of the simulation. Furthermore, field and collisional ionization processes are taken into account.

## Acknowledgements
We wish to acknowledge the operators and the scientific staff of the PHELIX facility for their support during the beamtime. We are particularly thankful to all collaborators of the Materials Research Department at GSI, who irradiated the polymer foils used to produce our targets. We would like to thank Prof. Peter Mulser and Dr. Ina Schubert for helpful discussions. The simulations were performed using HPC resources at TGCC/CCRT. We acknowledge PRACE for awarding us access to TGCC/Curie (Grant No. 2014112576).

### Author Contributions

P.N., M.E.T.-M. and D.K. designed the project, D.K., P.N. and B.B. conceived the experiment, L.B., L.M. and D.K. developed and produced the targets, D.K., F.G., B.B., F.W. and P.N. conducted the experiment, D.K. analyzed the laboratory data, M.L. and L.G. designed and performed the simulations, D.K., L.G., O.R. and P.N. interpreted the results and drafted the manuscript. All authors reviewed the manuscript.

### Additional Information

**Competing Interests:** The authors declare that they have no competing interests.

**Publisher's note:** Springer Nature remains neutral with regard to jurisdictional claims in published maps and institutional affiliations.